# Context-Based MEC Platform for Augmented-Reality Services in 5G Networks

Yue Wang, Tao Yu, and Kei Sakaguchi
Department of Electrical and Electronic Engineering, Tokyo Institute of Technology, Japan
{wang, yutao, sakaguchi}@mobile.ee.titech.ac.jp

*Abstract*—Augmented reality (AR) has drawn great attention in recent years. However, current AR devices have drawbacks, e.g., weak computation ability and large power consumption. To solve the problem, mobile edge computing (MEC) can be introduced as a key technology to offload data and computation from AR devices to MEC servers via 5th Generation Mobile Communication Technology (5G) networks. To this end, a context-based MEC platform for AR services in 5G networks is proposed in this paper. On the platform, MEC is employed as a data processing center while AR devices are simplified as universal input/output devices, which overcomes their limitations and achieves better user experience. Moreover, the proof-of-concept (PoC) hardware prototype of the platform, and two typical use cases providing AR services of navigation and face recognition respectively are implemented to demonstrate the feasibility and effectiveness of the platform. Finally, the performance of the platform is also numerically evaluated, and the results validate the system design and agree well with the design expectations.

## I. Introduction

As the latest generation of cellular mobile communication technologies, 5G has developed rapidly and been applied in many fields due to its impressive performance, such as ultra-high speed, ultra-low latency, and super reliability. It satisfies the requirements of those services featuring massive data exchange, real-time communication, and multi-scenarios. In recent years, plenty of 5G products have been rolled out for users, such as iPhone 12 and Lenovo Flex 5G. Beyond that, many other technologies and products are waiting for the arrival of 5G to achieve big innovations, one of which is AR, as well as its most typical application, Augmented reality (AR) glasses. As a kind of user terminal, AR glasses show a lot of advantages. For example, AR glasses can combine the real world and the virtual world to provide users with an immersive experience. Besides, AR glasses focus on the real-time interaction between users and surroundings, so they can serve as integrated tools to offer human-centric services.

However, existing AR glasses are still very niche products. Several reasons are hindering their widespread commercial application. Firstly, most AR glasses are large, heavy and their battery life is also short. This will seriously affect the user experience. Secondly, the computing ability is limited, so it is hard for AR glasses to perform complex functions. Such a weak processor will also bring high latency. The promising solution now is using cloud computing [1]. However, this will generate new problems of data transmission. Thirdly, up to now, many studies on AR are still focusing on application development. However, there are very few considerations in building an intact ecosystem for AR services. To this end, the technology of mobile edge computing (MEC) [2] in 5G networks can be a significant enabler, which will hopefully improve the status of AR and bring opportunities for the rapid development of AR glasses. The core idea of MEC is to offload computing data from terminal devices to the nearest edge. In this way, the powerful processor at the edge can support more computation-intensive task. Compared with cloud computing, the local area network of MEC will offer a reliable and low-latency connection between the edge server and AR glasses. In addition, data security and privacy can be guaranteed due to locality of data storage and data processing.

In recent years, many researchers have discussed the importance of 5G networks to AR technology and carried out evaluations on this topic [3][4][5]. Some studies also have begun to involve the application of MEC technology to AR. For example, Ren et al. proposed a hierarchical computation architecture for the implementation of the MEC-based AR framework. [6]. Jia et al. designed a model for supporting a massive multiplayer AR game by MEC [7]. Su et al. proposed a Mobile Augmented Reality (MAR) system deployed on a 5G edge testbed and showed end-to-end latency comparisons among 5G, LTE, and WiFi connections [8]. However, they are just limited to a certain single aspect, either only proving the feasibility of edge computing for AR improvement at the theoretical level, or only proposing to improve a certain AR application at the functional level, or only verifying the effectiveness of 5G in improving the real-time communication between AR devices and edge servers at the network level. In addition, there is very little literature studying on how to apply MEC and 5G into AR glasses for better services.

To this end, a novel platform for AR services using MEC technologies is designed in this paper, i.e., the context-based MEC platform, which overcomes the shortcomings associated with AR glasses and problems confronted by current computing schemes, and provides better user experience and effective AR development. Firstly, this platform gives a practical detailed answer about how to integrate MEC into AR in 5G, including the system structure, building blocks, communication strategy, etc. Secondly, this research does not focus on the improvement of one specific AR function or application scenario, but provides a general solution for AR service development under the platform. Based on this, the platform further enables users to get corresponding services based on their context scenarios.

The rest of the paper is structured as follows. Section 2 elaborates on the proposed platform for AR services in 5G networks. Section 3 describes two typical use cases as Proofs of Concept (PoC) from the perspectives of mechanism, workflow, and field experiments. Section 4 presents the numerical evaluation of the platform, including measurements using the PoC systems. Finally, section 5 concludes this paper.

## II. MEC Platform for AR Services

### A. System Overview

As mentioned above, MEC has been considered as a promising direction for AR in 5G networks. This section proposes a context-based application platform that integrates AR and MEC via 5G networks. As shown in Fig. 1, with this designed MEC platform, users can experience different AR services according to their corresponding scenarios. The scenarios are indicated by users' real-time context information about the physical environment and user behaviors, e.g., user location, surroundings, motion state. For example, when a user is in front of a gate, a navigation application would be turned on; when a user is in a dark office, a lighting-control system would be available. By using the platform, an ecosystem for AR services can be built up, where AR glass can get rid of the conventional service model and set up cooperation with the MEC server and other devices such as sensors and actuators. The MEC server will become the real AR service provider and data processor. Finally, AR glass can be simplified to a universal tool for collecting information and displaying 3D virtual objects.

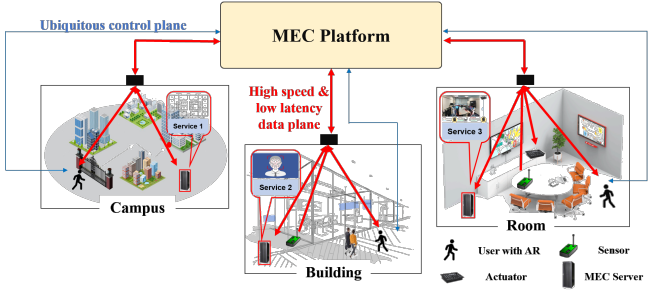

Fig. 1. Application Scenarios of Platform.

### B. System Architecture

Generally, edge computing uses resources of computing, storage, and network to provide services for users with low latency and high speed. The role of the MEC platform in this paper is to realize centralized control and management of these resources so that service providers can quickly develop and deploy AR applications. Referring to basic features of the standardized MEC framework [9], the architecture is designed as follows, which reflects the core idea of full-function MEC.

*1) MEC Platform*

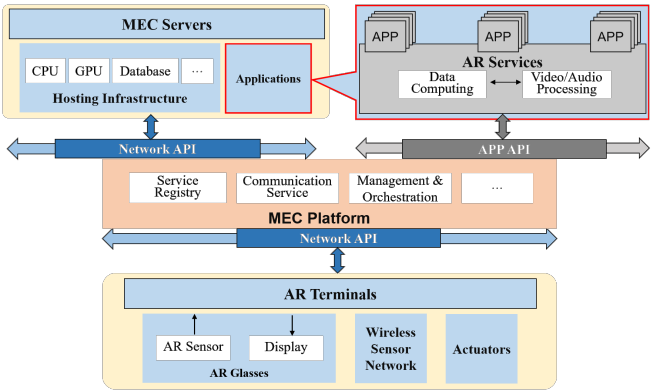

Fig. 2. System Operation Mechanism with MEC Platform.

As shown in Fig. 2, the MEC platform works as a middleware that configures a series of building blocks, which enable it to manage the overall system, distribute resources and AR applications over the application programming interfaces (API). On the one hand, the platform controls the connection between the AR terminals and edge servers based on the context of users by the control plane. On the other hand, the platform afterward invokes corresponding applications via application (APP) API and allows corresponding computation to be offloaded from glasses to MEC servers by the data plane via Network API. Under the MEC platform, there will be other three important parts involved in the system.

*2) AR Terminals*

The AR terminal part mainly includes AR glasses, as well as sensors and actuators. AR glasses are mainly responsible for input/output tasks. The input tasks include sensing the user information (location, gestures, voices, etc.) and transmitting the information to MEC servers. For some external information that cannot be collected by AR glasses, such as illuminance and temperature, a wireless sensor network (WSN) can be used to capture them. The output tasks include receiving and displaying analysis results from the MEC server. Similarly, actuators are used as auxiliaries to execute relevant operations based on the commands of users for external devices (e.g., lights, air conditioners).

*3) MEC Servers*

The MEC servers, which serve as the data processing center to perform computation-intensive tasks (e.g., deep learning, video processing). Once receiving data from AR terminals, MEC servers will process them, and then get the processing results. Finally, the results will be sent from the edge back to the AR glasses or actuators to execute. It is remarkable that the offloading strategy here from terminals to MEC servers is full offloading, which aims to offload entire computation data to process.

*4) 5G RAN Networks*

Fig. 3 shows the 5G Radio Access Network (RAN) as the data plane (D-Plane) for the platform, which consists of base stations deployed on the local area to realize the wireless connection among different components for data transmission. In addition, 4G networks will serve as the control plane (C-Plane) to manage the context information. For the networking method of 5G, there are many options. In our case, Non-Standalone (NSA) architecture is chosen, where 5G base stations will be deployed first and connect to the existing 4G core network. This network allows a large amount of computation to be offloaded from the terminal part to the MEC part with low latency.

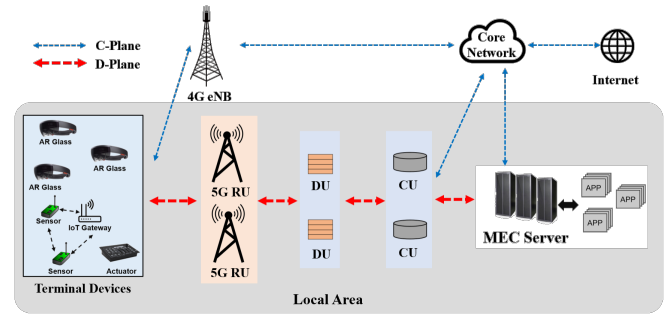

Fig. 3. Local 5G RAN Architecture for the Platform.

### C. Benefit of Platformization

Fully combing mobile edge computing and 5G networks with AR glasses, the designed platform has advantages as follows. 1) For users, the platform can greatly improve user experience, which includes, e.g., lower power consumption, smaller storage size, and lower physical weight and volume of AR glasses, by offloading the computation-cost task to edge servers via 5G to enable the low-latency, high-effectiveness and ultra-security AR services. 2) For developers, the platform can be a general solution for AR service development. With the platform that provides the required developing framework and core building blocks, developers can focus on developing the performance of services they want to implement without worrying about the underlying hardware, networks, system environment, etc. Consequently, developers can easily and efficiently design self-defined applications to realize desired AR services based on contexts.

### III. PROTOTYPE AND PROOF OF CONCEPT

In this section, prototype hardware is developed referring to the proposed platform. Based on it, two designed use cases with AR services, Navigation and Face Recognition, are implemented as its PoC.

#### A. Hardware and Software Implementation

For the prototype, the following hardware and software will be configured as instances. For the AR terminals, Microsoft HoloLens is chosen as the AR glass. The software used for HoloLens application development is Unity 3D with the software developer's kit named Mixed Reality Toolkit (MRTK). A PC with a graphics processing unit (GPU) serves as the MEC server. Since the MEC is the main part of the realization of the AR service functions, various software tools are used, e.g., OpenCV and TensorFlow. For the RAN part, due to limited hardware conditions, 5G networks are still not able to be deployed in the scenarios currently. Compared with other existing networks, in the local area, WiFi has similar performance to 5G in some respects, like high speed. Therefore, in the PoC part, we temporarily use the local WiFi instead as the wireless access network. Obviously, the platform can maximize its performance in 5G networks, but its features can also be reflected in WiFi. In addition, due to the limitation of HoloLens's hardware, it is impossible to realize completely full offloading. So instead, in the two use cases, the computation will be offloaded from HoloLens to the MEC server to the utmost extent.

#### B. Application Implementation of Use Cases

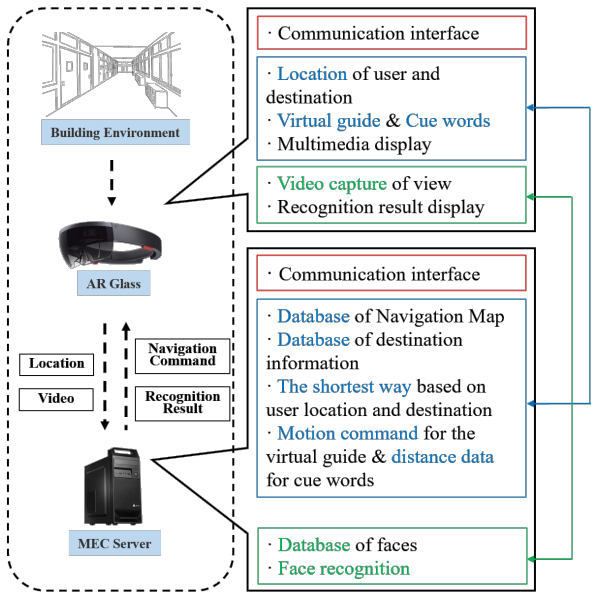

Fig. 4. Functional Mechanism of Use Cases.

The functional mechanism in Fig. 4 shows that the AR glass and the MEC server jointly complete two AR services by the platform, navigation (blue frame) and face recognition (green frame). The glass' and the server' communication interfaces (red frame) are connected during services.

Navigation service will lead users to the destination and provide destination information in real-time. For the AR glass, the input information is mainly location data of the user and destination. Its output task is to render the user interface (UI), including the virtual guide, cue words and the display frame. The MEC server stores databases of navigation map and destination information on the one hand, and calculates the shortest path based on location data and continually issues the navigation command to the AR glass on the other hand.

Face Recognition service can identify people's faces and display virtual labels in the glass view. The glass will capture the video by the glass camera in real-time as the input. Its output is displaying the recognition result. The MEC server will process videos or pictures received from AR glasses to recognize faces and get the results base on the face database.

#### C. Workflow and Field Experiments

##### 1) Use Case 1: Navigation

Fig. 5 presents the workflow of Navigation service. Firstly, the user should select the destination from the UI menu. Meanwhile, the AR glass detects the user's location. These two pieces of location information will be transmitted to the MEC server. Based on the location data, the server will get the shortest path instantly by the method of Dijkstra [10] and send back movement instructions to the AR glass. Then the virtual guide in the glass will lead the user to navigate along the shortest way until reaching the destination. Finally, the server will send the destination information image to the glass to display at the destination for the user.

Fig. 6 shows the AR diagrams of this use case in the field experiment respectively, i.e., 1) the third-person perspective; 2) the state of selecting destination; 3) the state of the virtual guide leading the user forward; 4) the scene of reaching the destination with the information image.

##### 2) Use Case 2: Face Recognition

The workflow of Face Recognition is shown in Fig. 7. Before the service, in the MEC server, the face database is trained to obtain the feature model. Once starting, the glass camera will capture videos or images and stream them to the MEC server in real-time. Then the server will process the video or images by the algorithm of Eigenface [11], including detecting faces, pretreating the frame with faces (e.g., resizing, graying), and extracting features. By matching the features with the trained feature model, the recognition result will be obtained and then sent back to the glass, where corresponding virtual labels are rendered at the face position in the view.

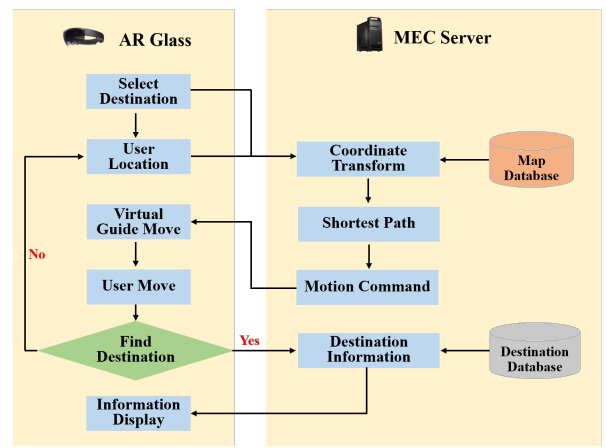

Fig. 5. Workflow of Navigation.

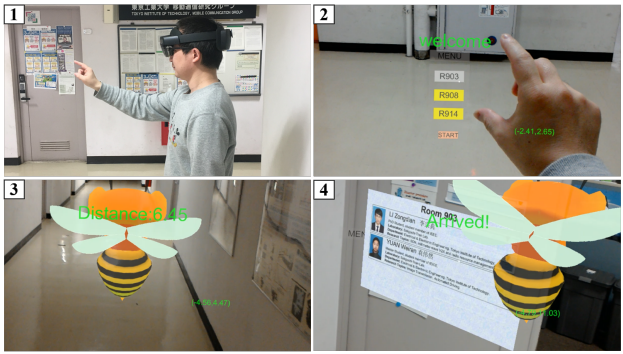

Fig. 6. Field Experiment of Navigation.

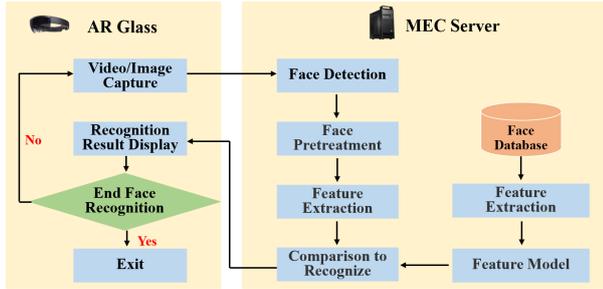

Fig. 7. Workflow of Navigation.

Fig. 8 shows the field experiment of Face Recognition, where recognizing multiple faces in real-time can be realized, i.e., 1) the third-person perspective; 2) the view of recognition result display.

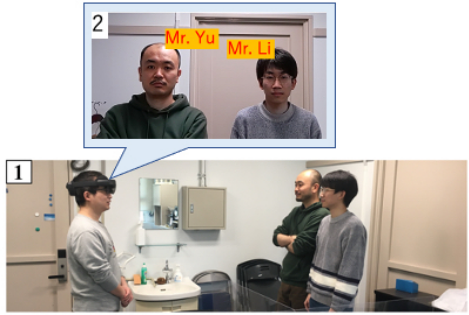

Fig. 8. Field Experiment of Face Recognition.

The successful implementation of the two use cases fully demonstrates feasibility and effectiveness of the MEC platform. On top of that, through different types of services in different scenarios, the use cases also prove the platform's scalability and versatility. The platform can be used as a general solution for the realization of AR applications.

## IV. EXPERIMENTS AND EVALUATION

In order to verify the performance advantages of the MEC platform over traditional methods, this section will evaluate performance by comparing the MEC platform with local AR device computing and cloud computing respectively.

### A. MEC Platform vs. Local Computing

The improvements in developing AR applications with the help of the MEC platform than with the AR glass itself can be reflected from the following three aspects.

*1) Storage.*

MEC allows part of data (like databases and models) to be deployed on MEC servers, which relieves the storage pressure of AR glasses, so that glasses can realize applications without requiring huge storage space. For example, in Use Case 2 of the PoC, deploying the face database on the MEC server will save 100M bytes storage space for the AR glass.

*2) Computing Ability*

With the platform, lots of computation will be offloaded from AR devices to MEC servers. To compare the data processing capabilities of the MEC server and the AR device, a set of contrast experiments is set up. The AR service selected here is face recognition as mentioned in Use Case 2. The workflows of the experimental groups, i.e., the MEC platform group and the local AR device group, are shown in Fig. 9 respectively. The face database, the face recognition model, and the face recognition algorithm adopted by the experimental groups are the same. The evaluation metric is the recognition time, which reflects the corresponding computing ability.

It can be seen from Fig. 10 that the MEC server takes much less time to complete the recognition, which means that its computing speed is faster and computation capacity is better. However, it should be noted that there is one additional step when using the MEC platform, i.e., video streaming. The test result of video stream latency was 300~400ms. Nevertheless, the total latency (the sum of computing time and video streaming latency) of the MEC platform is still slightly lower than that of the local AR device. There are many approaches to improve video streaming, like directly streaming uncompressed video or using 5G networks for faster transport. This can be a direction for future improvement.

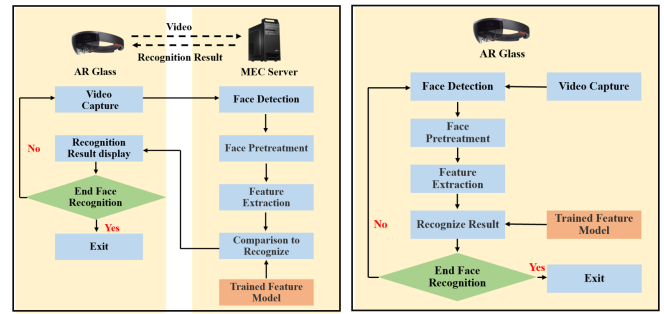

Fig. 9. Comparision of Edge Computing and Local Computing.

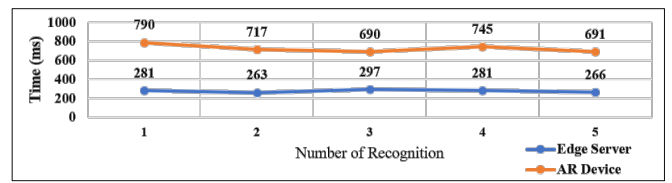

Fig. 10. Measurement Result of Computing Time.

*3) Contribution for the performance of AR device.*

This part will evaluate the contribution of the MEC platform for the performance of AR glasses during AR services. The contrast experiment setup is the same as that of computing ability evaluation. MEC platform group and local device group also keep the same. Evaluation metrics are the utilization of the central processing unit (CPU), GPU and power consumption of the AR glass.

Fig. 11 illustrates the average value of CPU, GPU utilization, and power consumption of the AR glass respectively during the face recognition service. It is apparent that with the MEC platform, the CPU utilization, GPU utilization and power consumption will be much reduced than the performance of the local device. On top of that, by comparing the utilization value of the MEC platform with the value of video streaming, it can be found that during services on the MEC platform, the utilization mainly attributes to the video streaming step. So, improving the performance of video streaming will enhance the advantages of the MEC platform.

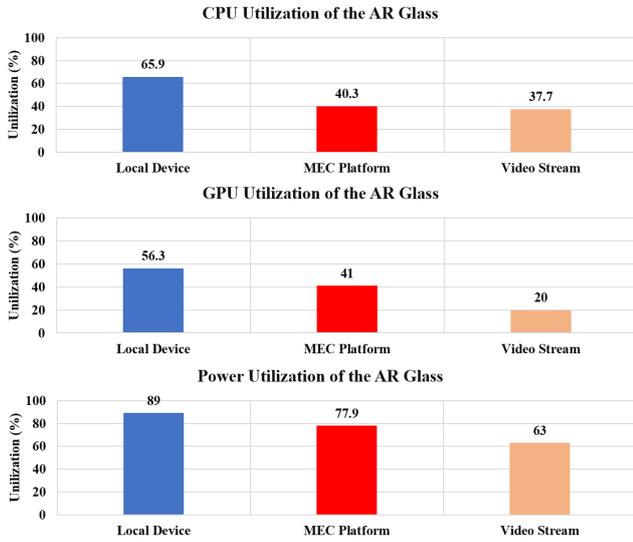

Fig. 11. CPU Utilization, GPU Utilization and Power Consumption.

### B. MEC Platform vs. Cloud Computing

Comparing with cloud computing, the MEC platform has better performance during AR services. In addition to ensuring data privacy, data security and network reliability, its merit can also be reflected from data communication.

For AR services, communication performance is supposed to be improved with the MEC platform. To this point, the E2E latency and throughput are evaluated with the contrast experiment as shown in Fig. 12. One experimental group is designed to simulate the cloud computing scheme, where the AR glass connects with the remote cloud server via World Interoperability for Microwave Access (WiMAX) as the blue route. The other experimental group is designed to simulate the edge computing scheme, where the AR glass and the MEC server are connected via local Wi-Fi as the black route.

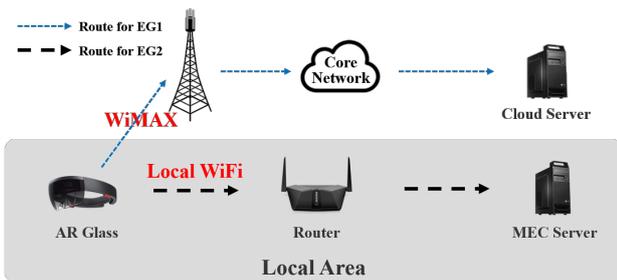

Fig. 12. Comparision of Edge Computing and Cloud Computing.

For the measurement of latency, the Transmission Control Protocol (TCP) packets are utilized to test instead of Internet Control Message Protocol (ICMP) packets. The measurement results in Fig. 13 shows that edge computing in the local area network has much lower latency than cloud computing performed remotely via the Internet.

For throughput measurement, HoloLens uploads the same file to the MEC server via local Wi-Fi and the cloud server via WiMAX respectively. The obtained results in Fig. 14 show the throughput value to the edge is much higher than that to the cloud, which proves the communication performance of the MEC platform is much excellent than that of cloud computing.

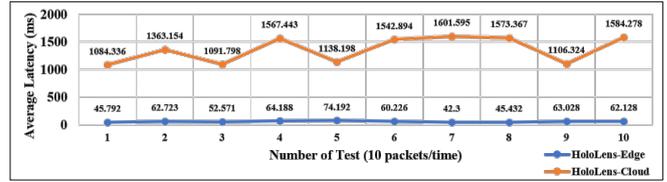

Fig. 13. Measurement Result of E2E latency.

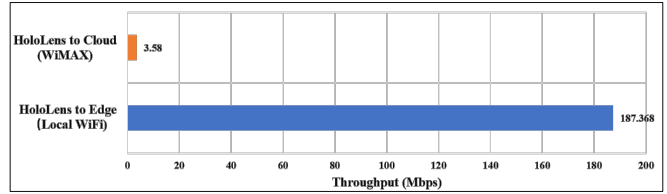

Fig. 14. Measurement Result of E2E throughput.

## V. CONCLUSION REMARKS

This paper briefly introduced AR and the technical capacities of MEC and 5G. On this basis, by combining these technologies, a novel MEC application platform for AR services was proposed, which not only improves the performance and user experience of services, but also provides the framework and building blocks for AR application development. Two practical use cases, Navigation and Face Recognition, were designed and implemented using the proposed MEC platform, which demonstrate the feasibility and effectiveness of the proposed MEC platform. Finally, this paper made quantitative evaluations of the MEC platform's performance by comparing the local computing scheme and the cloud computing scheme, and results showed its advantages in providing AR services.

## VI. ACKNOWLEDGEMENT

This work was supported by the NICT Beyond 5G R&D Promotion Project.